\documentclass[aip,reprint,a4paper] {revtex4-2}

\usepackage{upgreek}
\usepackage{amsfonts}
\usepackage{amstext}
\usepackage{mathtools}
\usepackage{amssymb}
\usepackage[dvipsnames]{xcolor}
\usepackage{amsmath}
\usepackage{graphicx}
\usepackage[normalem]{ulem}

\usepackage[urlcolor=blue]{hyperref}
\hypersetup{colorlinks=true,allcolors=blue}
\graphicspath{{./Figures/}}

\newsavebox\absfigbox
\newcount\absfiglines
\newlength\absfigwidth
\newlength\absfigsep
\newlength\absfigright   
\usepackage{microtype} 

\newcommand{\figExpPDZD}   {S1} 
\newcommand{\figPDZDCN}   {S2} 
\newcommand{\figMoSAIC}   {S3} 
\newcommand{\figDynCont}   {S4} 
\newcommand{\figPDZDfeatures} {S5} 
\newcommand{\figPDZZS}   {S6} 
\newcommand{\figPDZZL}   {S7} 
\newcommand{\figPDZZLct}   {S8} 
\newcommand{\tablePeakTimes}   {S1}

\begin{document}	
\author{Emanuel Dorbath}
\affiliation{Biomolecular Dynamics, Institute of Physics,
	University of Freiburg, 79104 Freiburg, Germany}
\author{Peter Hamm}
\affiliation{Department of Chemistry, University of Zurich, CH-8057
  Zurich, Switzerland}
\author{Gerhard Stock}
\affiliation{Biomolecular Dynamics, Institute of Physics,
	University of Freiburg, 79104 Freiburg, Germany}
\email{stock@physik.uni-freiburg.de}
\date{\today}
\title{Local molecular motions encode time-resolved infrared spectra of
proteins}

\begin{abstract}

\setlength{\emergencystretch}{2em}
\tolerance=1500
\pretolerance=0  

\setlength\absfigwidth{0.32\linewidth}
\setlength\absfigsep{0.035\linewidth}
\sbox\absfigbox{%
  \parbox[t]{\absfigwidth}{\centering
    \includegraphics[width=\linewidth]{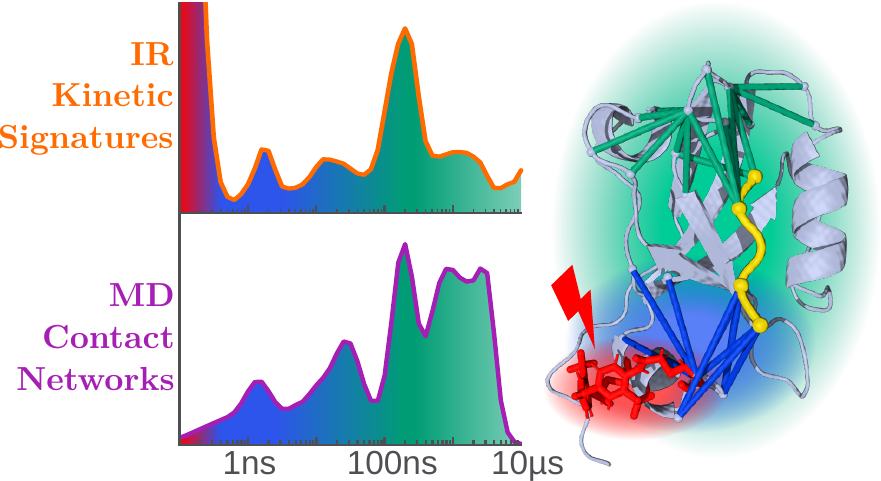}%
    }}

\absfiglines=\dimexpr\ht\absfigbox+\dp\absfigbox\relax
\divide\absfiglines by \baselineskip
\advance\absfiglines by 2

\hangindent=-\dimexpr\absfigwidth+\absfigsep\relax
\hangafter=-\absfiglines
\noindent
\rlap{\hspace*{\dimexpr\hsize-\leftskip-\rightskip
                            -\absfigwidth-\absfigright\relax}%
      \raisebox{-\height}[0pt][0pt]{\usebox\absfigbox}}%
Time-resolved infrared spectroscopy probes
\the\absfiglines\ 
  protein dynamics over timescales spanning more than ten orders of
  magnitude, yet the molecular motions underlying the observed kinetic
  signatures have remained elusive. Here we combine transient infrared
  spectroscopy with nonequilibrium molecular dynamics simulations to
  establish a direct connection between experimental relaxation times
  and local structural motions. Studying single-domain allosteric
  proteins, we find that inter-residue contact distances provide the
  structural representation that most faithfully reproduces the
  experimental dynamics. Correlation analysis identifies localized
  networks of coordinated contacts that mediate long-range coupling between
  secondary-structure elements. The characteristic timescales of these
  contact networks quantitatively match the experimentally observed
  relaxation processes, enabling each kinetic step to be assigned to a
  specific molecular motion. Applied to allosteric signal propagation
  in PDZ3 and photoinduced perturbation of the binding pocket of PDZ2
  or a ligand binding to PDZ2, this framework
  provides an atomistic picture of hierarchical protein relaxation and
  establishes a general framework for connecting transient infrared
  spectroscopy with the molecular mechanisms of protein dynamics.
\end{abstract}

\maketitle

%
%
\section*{Introduction}

Protein function emerges from structural dynamics spanning many orders
of magnitude in time, from local fluctuations to large-scale
conformational transitions involved in folding, binding, and
allosteric signaling.\cite{henzler-wildman_dynamic_2007,
  hensen_exploring_2012} Time-resolved infrared (IR) spectroscopy has
become a powerful tool for probing these nonequilibrium processes in
real time, \cite{bozovic_using_2022} complementing NMR spectroscopy,
which primarily probes equilibrium
dynamics. \cite{ishima_protein_2000, kay_new_2016} 
Transient IR spectroscopy is commonly used to study the response of
photoactive proteins on a wide range of
timescales. \cite{garczarek_functional_2006, 
kennis_ultrafast_2007, pfeifer_timeresolved_2009, kottke_grateful_2017, 
stritt_application_2020, hunt_biomolecular_2024} 
In particular, time-resolved IR studies of
photoswitchable PDZ domains and naturally photoactive proteins have
consistently revealed hierarchical relaxation extending from
picoseconds to milliseconds following
photoexcitation. \cite{buchli_kinetic_2013, bozovic_real-time_2020,
  bozovic_speed_2021,
  laptenok_infrared_2018,blankenburg_following_2019,
  buhrke_nanosecond_2020, ruf_molecular_2023} Although the observed
timescales can be explained by diffusion on a rugged free-energy
landscape, \cite{janke_universal_2025} linking them to specific
molecular motions has remained a major challenge.

Nonequilibrium molecular dynamics (MD) simulations provide atomistic
trajectories of these processes, \cite{buchenberg_time-resolved_2017,
  bozovic_real-time_2020,ali_allosteric_2024}
yet establishing a direct connection to transient IR experiments remains
challenging. The amide I signal reflects subtle changes in hundreds
of coupled backbone vibrations, making it difficult to attribute
spectral changes to specific structural events, both experimentally
and computationally.\cite{kobus_simulation_2011,baiz_vibrational_2020}
Rather than comparing spectra directly, we therefore focus on the
characteristic timescales of the nonequilibrium response.

According to transfer-operator theory and Markov state
models,\cite{prinz_markov_2011,noe_dynamical_2011} all observables of
a dynamical system share the same discrete relaxation timescales, differing
only in their amplitudes. Consequently we can compare the dynamical content
of experimental and simulated observables, which quantifies the amount
of dynamics occurring on each timescale.\cite{shaw_atomic-level_2010,
  stock_nonequilibrium_2018} The remaining challenge is therefore to identify
molecular observables whose dynamical content faithfully reproduces
that measured by transient IR spectroscopy.

Here we show that inter-residue contact distances
\cite{latzer_conformational_2008, ernst_contact-_2015,
  yao_establishing_2019} provide precisely such a molecular
representation. Combining with MoSAIC correlation
analysis,\cite{diez_correlation-based_2022, dorbath_contact_2026} we
identify localized contact networks that mediate long-range coupling between
secondary-structure elements. Across several photoswitchable PDZ
domains, the characteristic timescales of these contact clusters
quantitatively reproduce those measured by transient IR spectroscopy.
By combining previously reported transient IR measurements with
nonequilibrium MD simulations within a common dynamical framework, we
establish a quantitative connection between experimentally observed
relaxation timescales and the underlying dynamics of localized
molecular contact networks.

%
%
\section*{Results: Hierarchical dynamics of a photoswitchable PDZ3 domain}

PDZ domains are a canonical model system for allosteric communication,
\cite{cui_allostery_2008, tsai_unified_2014, thirumalai_symmetry_2019,
  wodak_allostery_2019} because of their well-characterized signaling
properties and compact
architecture. \cite{fuentes_ligand-dependent_2004, petit_hidden_2009,
  kumawat_hidden_2017, stevens_allosterism_2022} Their fold is highly
conserved, typically comprising two or three $\alpha$-helices and a
six‑strand $\beta$-sheet.
We first consider a photoswitchable PDZ3 domain that has been
characterized by both transient IR spectroscopy
\cite{bozovic_speed_2021} and nonequilibrium MD
simulations. \cite{ali_allosteric_2024} The system has served as a
minimal model for allosteric communication, in which a perturbation of
the short C-terminal $\alpha_3$-helix modulates the ligand-binding
affinity\cite{petit_hidden_2009}
(Fig.\,\ref{fig:PDZ3}a). In the experiment of Bozovic et
al.,\cite{bozovic_speed_2021} the $\alpha$-helical content of
$\alpha_3$ is controlled by covalently linking two residues with an
azobenzene moiety. In the \textit{cis} configuration the azobenzene
matches the native end‑to‑end distance of the helix, thereby
preserving its secondary structure. Photoisomerization to the
\textit{trans} state elongates the linker, enforcing partial unfolding
of $\alpha_3$ and thereby inducing the allosteric transition.

%
%
\subsection*{Experimental results}

\begin{figure}[ht!]
\includegraphics[width=.46\textwidth]{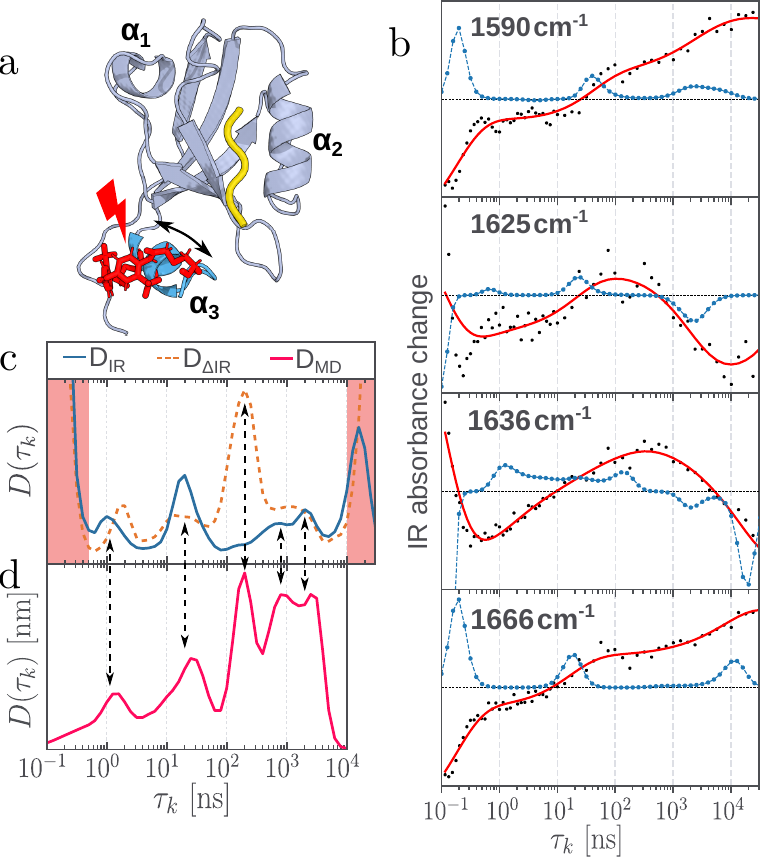}
\caption{
  Time-resolved IR spectroscopy of a photoswitchable PDZ3
  domain. (a) Molecular structure including
  photoswitch (red) and ligand (yellow). Upon
  \textit{cis}-to-\textit{trans} photoisomerization, the
  $\alpha_3$-helix detaches from the protein.  (b) Representative
  examples of the time-resolved IR response $I(t,\omega)$ of
  ligand-bound PDZ3, measured at four probe frequencies
  $\omega$. The data are shown as black dots, the fits according to
  Eq.\,(\ref{eq:TSA}) as red line and the resulting timescale spectra
  as blue line. (c) Dynamical content (in arbitrary units) of the IR
  response of ligand-bound PDZ3, $D_{\rm IR}$, and of the difference
  of ligand-bound and unbound signals, $D_{\Delta {\rm IR}}$. Data at
  very short and very long timescales are beyond the scope of our MD
  study and are therefore shaded in red. All experimental data are
  adapted from Ref.~\citenum{bozovic_speed_2021}, using a
    modified regularization of the timescale analysis to facilitate comparison with
    the MD results. (d) Dynamical content $D_{\rm MD}$ calculated from MD simulations.
    \label{fig:PDZ3}}
\end{figure}

Transient IR spectroscopy reveals highly heterogeneous kinetics across
the amide I band. Representative time traces $I(t,\omega)$ measured at
four probe frequencies are shown in Fig.\,\ref{fig:PDZ3}b as a
function of logarithmic time.\cite{bozovic_speed_2021} While the
signals at $\omega = 1590$ and 1666\,cm$^{-1}$ increase monotonically,
we also find signals with one or two extrema, most notably a broad
maximum around 100\,ns.

To characterize the multiscale dynamics, we decompose each transient
IR signal into a superposition of exponential
relaxations,\cite{stock_nonequilibrium_2018} 
\begin{equation}\label{eq:TSA}
I(t,\omega_j) \approx \sum_{k} a_{k j} \,e^{-t/ \tau_{k}} ,
\end{equation}
where the time constants $\tau_k$ are distributed uniformly on a
logarithmic scale (typically 10 per decade). The amplitudes $a_{kj}$
are determined by maximum-entropy
regularization,\cite{lorenz-fonfria_transformation_2006}
and the resulting fits are shown as red curves in Fig.\,\ref{fig:PDZ3}b. The
resulting amplitudes $a(\omega_j,\tau_k)=a_{kj}$ define the
timescale spectra (blue), where positive and negative peaks denote
kinetic components that increase or decrease the absorption,
respectively. Details of all timescale analyses are given in the SI.

To condense the timescale information contained in the complete
transient IR data set, we combine the timescale spectra obtained at
all probe frequencies into a single \emph{dynamical 
  content},\cite{shaw_atomic-level_2010,stock_nonequilibrium_2018} 
\begin{equation}\label{eq:DynCont}
D(\tau_k) = \sqrt{ \sum_j |a_{k j}|^2 },
\end{equation}
which measures the total dynamical activity occurring on the timescale
$\tau_k$ while preventing cancellation of positive and negative
spectral contributions. 
The dynamical content exhibits approximately
one prominent peak per decade (Fig.\,\ref{fig:PDZ3}c). 
While this peak pattern appears to be a rather universal feature of
the nonequilibrium response of compact proteins, \cite{janke_universal_2025} the present analysis provides an atomistic interpretation of the individual peaks by
assigning them to localized contact networks and specific molecular
rearrangements.

To isolate the effect of ligand binding on the nonequilibrium
response,\cite{bozovic_speed_2021} Bozovic et al. also measured
transient IR signals of ligand-free PDZ3. Although the dynamical
contents of the bound and unbound proteins differ only subtly
(Fig.\,\figExpPDZD), the dynamical content of their difference
exhibits a pronounced peak near 200 ns (Fig.\,\ref{fig:PDZ3}c).  This
peak therefore reports the time required for the perturbation to
propagate from the photoswitch to the ligand, providing a direct
experimental measure of allosteric signaling in a single PDZ
domain. \cite{bozovic_speed_2021}

%
%
\subsection*{MD modeling}

To obtain an atomistic description of the allosteric transition, Ali
et al.\cite{ali_allosteric_2024} performed
extensive nonequilibrium all-atom MD simulations of the
photoswitchable PDZ3 domain. Following equilibrium sampling,
photoisomerization of the azobenzene linker was modeled using a
potential-energy-surface switching
approach,\cite{nguyen_photoinduced_2006} yielding 112
nonequilibrium trajectories ($90 \!\times\! 1\, \mu$s and
$22 \!\times\! 10\, \mu$s; $\sim 15\times10^6$ MD frames). From these
trajectories we extracted 403 inter-residue heavy-atom contacts,
defined by a minimum heavy-atom distance below 4.5\,\AA\ and a
population exceeding 10\,\%.\cite{ernst_contact-_2015,
  yao_establishing_2019}

\begin{figure}[ht!]
\includegraphics[width=.46\textwidth]{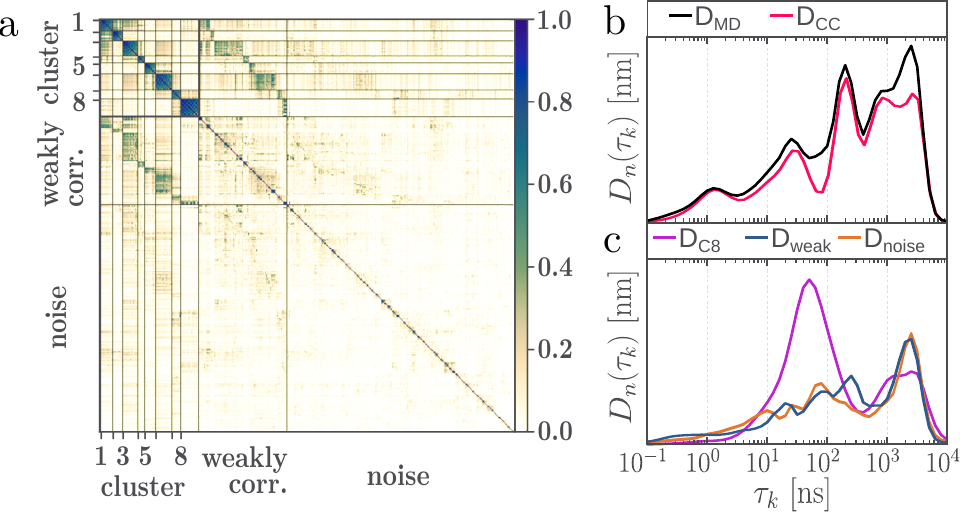}
\caption{
  Nonequilibrium MD analysis of the allosteric transition in
  PDZ3. \cite{ali_allosteric_2024} (a) Block-diagonal MoSAIC
  correlation matrix of inter-residue contact distances (with
  resolution parameter $\gamma=0.5$), revealing eight clusters of
  correlated contacts together with weakly correlated and uncorrelated
  contacts. (b, c) Timescale analysis of nonequilibrium MD
  trajectories yields dynamical contents $D_\alpha(\tau_k)$ for
  different subsets $\alpha$ of contact distances. $D_{\rm MD}$
  includes all contact distances except those involving the N
  terminus; $D_{\rm CC}$ comprises contacts from clusters C1–C7;
  $D_{\rm C8}$ contains only contacts from cluster C8; $D_{\rm weak}$
  includes weakly correlated contacts; and $D_{\rm noise}$ consists of
  uncorrelated contacts.}
  \label{fig:PDZ3MD1}
\end{figure}

Only a subset of inter-residue contacts participates in the allosteric
transition. To identify these contacts, we applied the MoSAIC
framework.\cite{diez_correlation-based_2022} The correlation matrix of
all contact distances was reordered using Leiden community detection into
an approximately block-diagonal form (Fig.\,\ref{fig:PDZ3MD1}a),
revealing eight strongly correlated contact clusters together with
weakly correlated and uncorrelated contacts. Clusters C1–C7
correspond to the previously identified functional contact networks,
\cite{ali_allosteric_2024} whereas C8 reflects
fluctuations of the flexible N terminus.
We first focus on the seven functional clusters, comprising 73 contact
distances. Following the smoothing of the time traces via Gaussian
filtering, \cite{sartore_lost_2026} we performed a timescale analysis
on their time‑dependent average distances and extracted the associated
dynamical content, $D_{\rm CC}$. The dynamical content $D_{\rm CC}$
exhibits peaks near $1$\,ns, 20\,ns, 200\,ns, 800\,ns, and $2\,\mu$s
(Fig.\,\ref{fig:PDZ3MD1}b), remarkably similar to the experimental
peaks observed in Fig.\,\ref{fig:PDZ3}c.
(See Figs.\,\figMoSAIC{} and \figDynCont{} for validations of MoSAIC
clustering and dynamical content calculations, and
Tab.\,\tablePeakTimes{} for a list of all peak times.)

Transient IR spectroscopy could in principle also be sensitive to the
remaining contacts. Cluster C8 comprises 17 contacts involving the
highly flexible N terminus and exhibits a pronounced peak near 50\,ns
(Fig.\,\ref{fig:PDZ3MD1}c), which we expect to disappear with improved
MD sampling (see Fig.\,\figPDZDCN). The uncorrelated contacts produce
only a broad, low-amplitude background extending from 2\,ns to
200\,ns, followed by a weak feature near $2\,\mu$s, reflecting the
cumulative contribution of many stable contacts.  Weakly correlated
contacts display essentially the same characteristic timescales as
$D_{\rm CC}$, but with substantially smaller amplitudes
(Fig.\,\ref{fig:PDZ3MD1}c).

For direct comparison with the experimental data, we therefore include
all contacts except those involving the N terminus. The resulting
dynamical content, $D_{\rm MD}$, closely matches $D_{\rm CC}$
(Fig.\,\ref{fig:PDZ3MD1}b). This close agreement suggests that the
experimentally observed IR kinetics is dominated by the dynamics of
the functional contact clusters, a correspondence that we examine
directly in the following section.

%
%
\subsection*{Comparison of IR and MD response} \label{sec:Comp}

We now compare the MD-derived dynamical content $D_{\rm MD}$ with the
experimental dynamical content $D_{\rm IR}$ (Fig.\,\ref{fig:PDZ3}b).
Two experimental features should be excluded from the
comparison. First, $D_{\rm IR}$ exhibits a strong signal near 0.1\,ns,
arising from rapid heating following the \emph{cis}-to-\emph{trans}
photoisomerization of azobenzene, \cite{hamm_vibrational_1997} which
is unrelated to the structural relaxation considered here. Second, the
experimental relaxation near $20\,\mu$s lies beyond the $10\,\mu$s
time window accessible to the simulations. We therefore restrict the
comparison to timescales between 0.5\,ns to $10\,\mu$s.

The resulting comparison of $D_{\rm IR}$ and $D_{\rm MD}$ in
Figs.\,\ref{fig:PDZ3}b,c shows a striking agreement in the locations
of the characteristic timescales. Although the absolute peak heights
differ, the simulations reproduce the experimental features at
approximately 1\,ns, 20\,ns, 800\,ns and $2\,\mu$s with high fidelity.
The strong peak at 200\,ns that appears in $D_{\rm MD}$, but is barely
visible in $D_{\rm IR}$ of ligand-bound PDZ3, becomes evident in the
dynamical content $D_{\Delta {\rm IR}}$ associated with the difference
of the bound and unbound IR signals. Thus every experimentally
observed relaxation process can be matched to a corresponding feature
in the MD-derived dynamical content. 

%
%
\subsection*{Contact cluster interpretation of the IR response}

\begin{figure}[ht!]
\includegraphics[width=.35\textwidth]{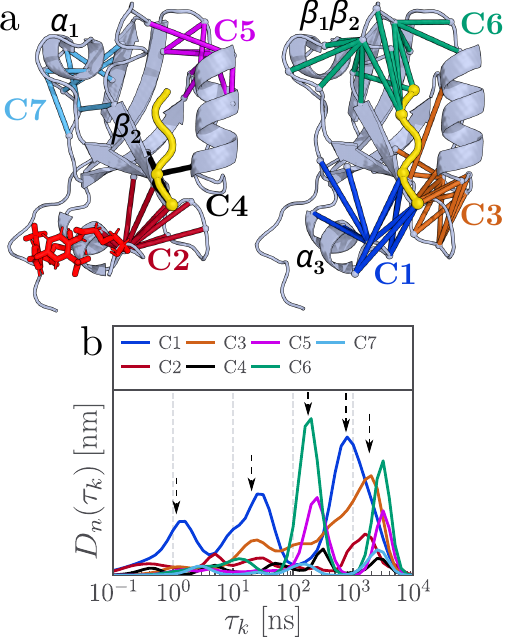}
\caption{
  Contact cluster analysis of the nonequilibrium response of PDZ3. (a)
  Molecular structure of PDZ3, illustrating the contact distances
  associated with clusters C1 to C7 by colored lines. (b) Dynamical
  contents of the individual clusters. Arrows indicate the
  timescales shown in Fig.\,\ref{fig:PDZ3}b,c.} \label{fig:PDZ3MD2}
\end{figure}

The agreement between the experimental and MD-derived dynamical
contents (Fig.\,\ref{fig:PDZ3}b,c) demonstrates that the transient IR
response directly reports the dynamics of localized contact clusters.
These clusters consist primarily of tertiary contacts linking rigid
secondary-structure elements and therefore define the flexible joints
that mediate large-scale conformational changes
\cite{papaleo_role_2016} (Fig.\,\ref{fig:PDZ3MD2}a)

To characterize the nonequilibrium response of the individual contact
clusters, we analyze their dynamical content $D_n(\tau_k)$ computed
from the contact distances assigned to cluster $n$. These quantities
are related to the overall dynamical content of the contact-cluster
representation through
$D_{\rm CC}(\tau_k) = \sqrt{\sum_{n} D_n(\tau_k)^2}$ The
cluster-resolved dynamical contents (Fig.\,\ref{fig:PDZ3MD2}b) reveal
that C1, C3, and C6 dominate the response. The initially excited
cluster C1 exhibits peaks near 1, 30, and 800\,ns, whereas the 200\,ns
feature arises mainly from C6, with contributions from C4 and C5. At
2-3\,$\mu$s, nearly all clusters contribute, led by C6, C3, and C5.

Structural analysis \cite{ali_allosteric_2024} assigns the 1 and
20\,ns features to rupture of the initial contacts followed by
detachment of the $\alpha_3$-helix. The prominent 200\,ns response
occurs in the C6 cluster, located in the distant $\beta_1$--$\beta_2$
loop, indicating that the allosteric perturbation initiated in the
$\alpha_3$ region propagates across the PDZ domain on this
timescale. This long-range coupling is mediated by the ligand, which
connects contacts with $\alpha_3$ to those with $\beta_1$--$\beta_2$
and thereby transmits the structural perturbation through successive
contact-network rearrangements.\cite{ali_allosteric_2024} The close
agreement between the simulated and experimental timescales provides
strong support for assigning the 200\,ns process to allosteric signal
propagation in PDZ3. \cite{bozovic_speed_2021} Capturing this
spatiotemporal sequence of structural changes requires nonequilibrium
all-atom MD simulations.

The MD simulations further reveal a previously unrecognized
realignment of the $\alpha_3$-helix at $\sim$800\,ns through
cooperative formation of new ligand and protein
contacts.\cite{ali_allosteric_2024} Finally, the concerted relaxation
of nearly all clusters at 2-3\,$\mu$s indicates that the allosteric
transition follows an ``order-disorder-order'' mechanism, in which
transient local unfolding precedes formation of the final equilibrium
state.\cite{buchenberg_time-resolved_2017,stock_nonequilibrium_2018}

%
%
\subsection*{Performance of alternative feature sets}

The success of contact distances naturally raises the question of
whether other structural descriptors perform equally
well. Replacing heavy-atom contact distances by C$\alpha$ contacts
($d_{\rm C\alpha}<0.8$\,nm) yields similar contact
clusters,\cite{dorbath_contact_2026} indicating that C$\alpha$
distances provide a useful structural approximation. The
corresponding dynamical content, $D_{\rm C\alpha}$
(Fig.~\figPDZDfeatures), reproduces the main peaks at 20\,ns, 200\,ns,
and 2\,$\mu$s but misses the 800\,ns feature and underestimates the
early-time response. These missing features originate from detachment and
realignment of the $\alpha_3$-helix, which involve reorganization of
long side chains not captured by C$_\alpha$ distances.

Using all pairwise C$_\alpha$ distances despite their strong
redundancy, produces a nearly
identical dynamical content (Fig.~\figPDZDfeatures). In contrast,
backbone dihedral angles yield only a weak, featureless response and
fail to resolve the dominant timescales. These findings indicate that
the allosteric transition is encoded primarily in changes of
inter-residue contacts and side-chain packing rather than backbone
$(\phi,\psi)$ rearrangements.

%
%
\section*{Photoswitchable PDZ2 constructs} 

Having established the correspondence between transient IR
spectroscopy and contact-cluster dynamics for PDZ3, we now ask whether
the same framework applies to structurally distinct PDZ2
constructs. \cite{buchli_kinetic_2013, buchenberg_time-resolved_2017,
  bozovic_real-time_2020} Unlike PDZ3, these constructs lack the
$\alpha_3$-helix and probe ligand binding and release through direct
perturbations of the binding pocket.

\begin{figure}[ht!]
\includegraphics[width=.46\textwidth]{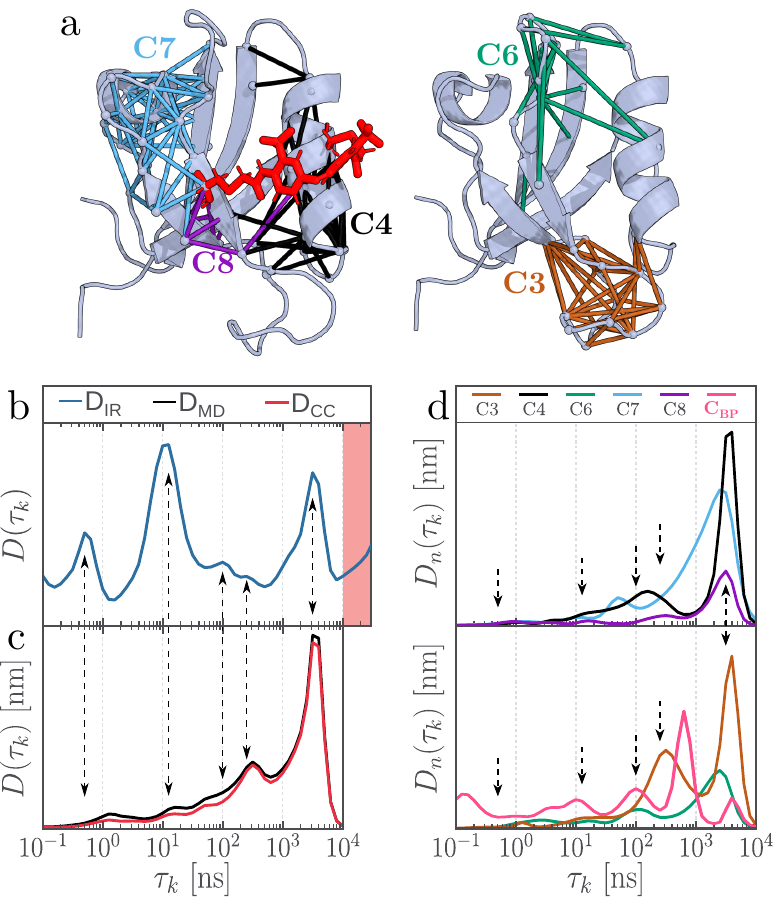}
\caption{
  PDZ2S featuring a photoswitch
  (red) across its binding pocket. (a) Structure and contact
  clusters. (b-d) Dynamical contents of PDZ2S,
  obtained from IR spectroscopy ($D_{\rm IR}$), from all contact
  distances except terminal distances ($D_{\rm MD}$), from the five
  functional contact clusters ($D_{\rm CC}$), from the individual
  clusters ($D_n$), and from the C$_\alpha$-distances bridging the
  binding pocket ($D_{\rm BP}$).}
\label{fig:PDZ2S}
\end{figure}

%
%
\subsection*{PDZ2S}

In PDZ2S, an azobenzene linker connects residue Azo22 in the
$\beta_2$-sheet to Azo77 in the
$\alpha_2$-helix.\cite{buchli_kinetic_2013} Upon
\textit{cis}-to-\textit{trans} photoisomerization, linker elongation
drives the binding pocket from a conformation resembling the unbound
state toward the bound state (Fig.\ \ref{fig:PDZ2S}a).
Buchli et al.\cite{buchli_kinetic_2013}
conducted time-resolved IR spectroscopy of the photoswitchable protein
PDZ2S and observed surprisingly complex dynamics. Applying the
timescale analysis to all IR transients yields the dynamical content $D_{\rm IR}$
of the experiment was obtained (Fig.\,\ref{fig:PDZ2S}b.) Similar to
the results for PDZ3, the dynamical content of PDZ2S exhibits
roughly one peak per time decade from $\sim 0.1\,$ns to
$20\,\mu$s. This similarity is remarkable because the global perturbation of
PDZ2S is expected to result in a quite different molecular response
than the local perturbation of PDZ3.

To rationalize the experimental observations, we apply the same
workflow to the nonequilibrium trajectories ($14\!\times\!10\,\mu$s)
of Buchenberg et al.\cite{buchenberg_time-resolved_2017} From 330
heavy-atom contact distances, MoSAIC clustering identified five
functional contact clusters, two terminal clusters, and a noise-like
background (Fig.\, \figPDZZS). As the terminal clusters
contribute only weakly, we excluded them. The resulting dynamical
content, $D_{\rm MD}$, exhibits dominant peaks at $\sim$300\,ns and
4\,$\mu$s, with weaker features at 1 and 10\,ns (Fig.\
\ref{fig:PDZ2S}c). These peaks closely match the experimental profile
$D_{\rm IR}$, except for a stronger experimental response at short
times.

As in PDZ3, $D_{\rm MD}$ closely follows the combined response of the
functional contact clusters, $D_{\rm CC}$, again allowing the
experimental kinetics to be interpreted structurally. Clusters C3, C6,
and C7 are conserved between PDZ2S and PDZ3, whereas the opening of
the binding pocket gives rise to cluster C4 around the
$\alpha_2$-helix and cluster C8 linking the $\beta_2$ and $\beta_3$
strands (Fig.\,\ref{fig:PDZ2S}a). Cluster-resolved dynamical contents
assign the 10\,ns feature to photoinduced opening of the binding
pocket (mainly C8), the 1\,ns response with local rearrangements near
the photoswitch (C4, C6, and C8), the 300\,ns peak with reorganization
of C3 and C4, and the 4\,$\mu$s peak with relaxation into the fully
open state. Smaller responses from the distant clusters C6 and C7
demonstrate that the structural perturbation propagates throughout the
protein.

Using C$\alpha$ contacts or all C$\alpha$ distances yields similar
dynamical contents but enhances the short-time response
(Fig.\,\figPDZZS). In particular, C$_\alpha$ distances spanning the
binding pocket display pronounced peaks at early times, supporting the
assignment of the first experimental features to binding-pocket
opening (Fig.\,\ref{fig:PDZ2S}d). Their enhanced amplitudes suggest
that transient IR spectroscopy is also sensitive to structural motions
beyond inter-residue contact rearrangements alone.
%
%
\subsection*{PDZ2L}

\begin{figure}[ht!]
\includegraphics[width=.46\textwidth]{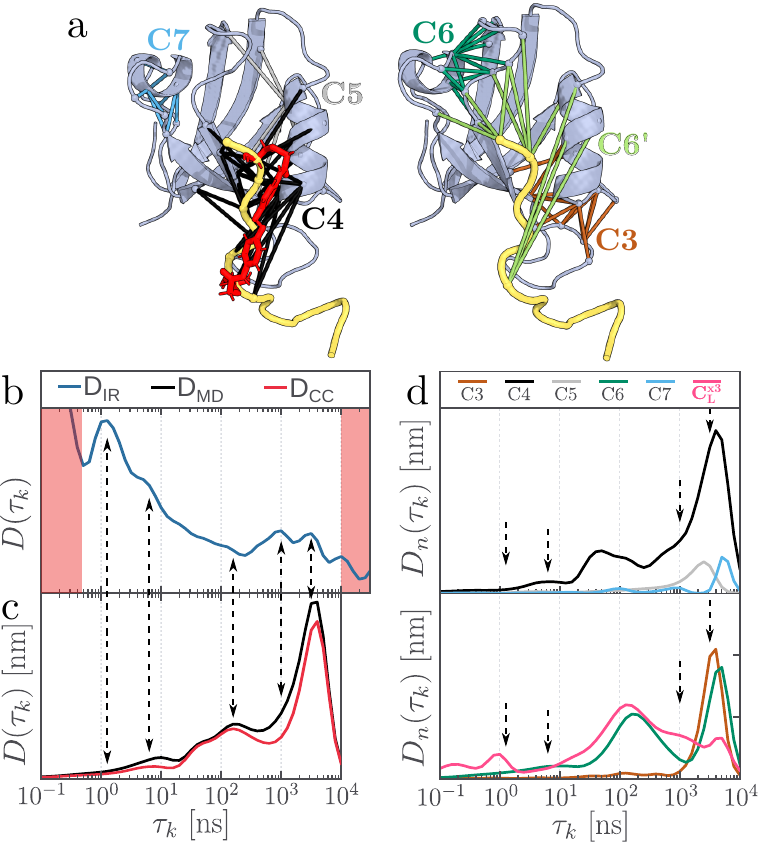}
\caption{
  PDZ2L containing a photoswitchable ligand. (a) Structure and contact
  clusters. (b-d) Dynamical contents of PDZ2L, as obtained from IR
  spectroscopy ($D_{\rm IR}$), from all contact distances except
  terminal distances ($D_{\rm MD}$), from the six functional contact
  clusters ($D_{\rm CC}$), from the individual clusters ($D_n$), and
  from intra-ligand C$_\alpha$-distances ($D_{\rm Lig}$).}
\label{fig:PDZ2L}
\end{figure}

In the PDZ2L construct, the photoswitch is attached to the ligand
rather than the protein,\cite{bozovic_real-time_2020} so {\em
  trans}$\rightarrow${\em cis} photoisomerization compresses the
ligand, destabilizes its interactions with the binding pocket, and
initiates unbinding (Fig.\,\ref{fig:PDZ2L}a). Because the perturbation
acts directly on the ligand rather than on the protein scaffold, PDZ2L
provides a more realistic model than PDZ2S.

The experimental dynamical content, 
$D_{\rm IR}$, differs markedly from that of PDZ3 and PDZ2S: following
an initial peak near 1\,ns, it decreases continuously until
$\sim$200\,ns and exhibits only weak features at 1, 3.5, and
10\,$\mu$s (Fig.\,\ref{fig:PDZ2L}b).
Applying the same analysis to extensive nonequilibrium MD simulations
\cite{bozovic_real-time_2020,dorbath_contact_2026} ($80\times 1\,\mu$s
and $19\times 10\,\mu$s) identified six functional contact clusters
together with terminal, ligand-tail, weakly correlated, and noise-like
motions (Fig.\,\figPDZZL). After excluding the highly mobile termini
and unbound ligand tail, the MD dynamical content, $D_{\rm MD}$, rises
from $\sim$1\,ns to a dominant peak near 3\,$\mu$s, with weaker
features at 10 and 150\,ns. The characteristic timescales agree well
with experiment, although their amplitudes show opposite trends.

As $D_{\rm MD}$ is again dominated by the functional contact clusters,
the experimental response can again be interpreted in terms of their
dynamics. Besides the conserved PDZ clusters, ligand unbinding gives
rise to two binding-pocket clusters (C4 and C6') linking ligand and
protein (Fig.\ \ref{fig:PDZ2L}a). These binding-pocket clusters respond first,
producing broad features between 8 and 150\,ns, whereas the dominant
$\sim$4\,$\mu$s peak reflects concerted rearrangement of nearly all
clusters after the ligand has left its binding pose.

Alternative descriptors based on C$\alpha$ contacts or distances yield
similar dynamical contents (Fig.\,\figPDZZL). Intra-ligand C$\alpha$
distances reveal an additional 1\,ns peak, indicating that rapid
ligand rearrangements contribute to the early response, although they
cannot fully account for the experimental amplitude.

Although the simulations of ligand rebinding were insufficiently
sampled for a quantitative comparison,\cite{dorbath_contact_2026}
their dynamical content nevertheless reproduces the experimentally
observed timescales remarkably well, with corresponding features near
0.6\,ns, 50\,ns, 400\,ns, and 3\,$\mu$s, see Fig.~\figPDZZLct{}.

%
%
\section*{Discussion}

We have established a general framework that links transient IR spectroscopy
to atomistic protein motions through nonequilibrium MD simulations.
By analyzing experimental and simulated time traces with
multi-exponential functions, we have constructed the dynamical content
of the protein response, \cite{shaw_atomic-level_2010,
  stock_nonequilibrium_2018} defined in Eq.\,(\ref{eq:DynCont}) as the
distribution of dynamical activity across timescales. Within the
framework of Markov state model theory,\cite{prinz_markov_2011,
  noe_dynamical_2011} the characteristic timescales extracted from
experimental and MD-derived dynamical contents can be compared
directly, even though their amplitudes generally may differ.
Thus, the amplitude differences between experiment and computation
reflect the different sensitivities of the IR and MD observables to
the underlying dynamics, rather than differences in the underlying
relaxation timescales.

Among the structural descriptors examined, heavy-atom contact
distances provided the most faithful description of the experimental
timescales. Unlike $(\phi,\psi)$ backbone dihedral angles or
C$_\alpha$-distances, contact distances directly describe the
formation and rupture of local interactions that stabilize protein
structure. \cite{ernst_contact-_2015, yao_establishing_2019,
  nagel_selecting_2023}
MoSAIC correlation analysis\cite{diez_correlation-based_2022} further
revealed that only a small number of localized contact clusters dominate
the experimentally relevant dynamics, whereas weakly correlated and
noisy contacts contribute mainly a diffuse background.

Notably, changes in polar contacts, such as the formation or rupture
of salt bridges, typically induce substantial changes in residue
electrostatic energies.\cite{kumawat_hidden_2017} These changes alter
the electric field throughout the protein and, in particular, around
the backbone C=O groups that determine the amide I
frequencies.\cite{baiz_vibrational_2020} Other mechanisms known to
influence the amide I frequencies, including fluctuations of the
backbone $(\phi,\psi)$ dihedral angles and changes in hydrogen bonding
to the backbone C=O and N-H groups, are expected to play only a minor
role in structurally stable proteins such as PDZ domains. The
electrostatic changes associated with contact rearrangements therefore
emerge as the dominant mechanism linking protein structural dynamics
to the transient amide I response.  It has been suggested before that
long-range electrostatics interactions play an important role in
allosteric communication. \cite{banerjee-ghosh_longrange_2020a} Our
results thus revise the molecular interpretation of time-resolved IR
spectroscopy by showing that its relaxation times directly report the
dynamics of localized contact networks that mediate conformational
transitions.

Across all three systems we found excellent agreement between
experiment and simulation. For PDZ3, the analysis revealed that the
experimental timescales at 1, 10\,ns and 800\,ns reflect the
initial detachment and subsequent realignment of the $\alpha_3$-helix
relative to the protein core. In contrast, the timescales at 200\,ns
and 3\,$\mu$s account for the allosteric response of the distal
$\beta_1$--$\beta_2$ loop. These findings provide a mechanistic
interpretation of the observed dynamical content, explaining the
approximately one-peak-per-decade as the signature
of a hierarchical relaxation process composed of distinct,
structurally resolved steps.
Moreover, the predicted cooperative rearrangements of contact networks
provide experimentally testable hypotheses that could be probed
directly by time-resolved X-ray structural
methods. \cite{boutet_high_2012, barends_direct_2015}

Applying the same MD analysis workflow, we were able to reproduce
virtually all experimentally observed timescales in both PDZ2
systems. Capturing the short-time response required the inclusion of
C$_\alpha$-distances reporting on the photoswitch region, as only a
limited number of contact distances are present in this part of the
structure.
While the characteristic timescales are reproduced almost
quantitatively, the amplitudes differ between
experiment and simulation. For both PDZ2S and, even
more prominently, PDZ2L, the IR-derived dynamical contents are
dominated by large amplitudes at short timescales and comparatively
weak contributions at long times. In contrast, the MD-derived
dynamical contents exhibit the opposite trend, with the largest
amplitudes occurring on longer timescales.

The latter behavior can be rationalized, at least in part, by the
diffusive nature of conformational transitions in
proteins.\cite{zwanzig_diffusion_1988, neusius_subdiffusion_2008,
  janke_universal_2025} Within this framework, structural relaxation
proceeds diffusively, so that the average change of many inter-residue
distances grows approximately as a power law, $\propto t^{\alpha}$,
before eventually saturating in the new equilibrium state.  Dorbath et
al.\cite{dorbath_contact_2026} found this behavior for the majority of
contact distances in PDZ domains at long times, showing that the exponent $\alpha$
typically ranges between 0.3 (anomalous or subdiffusive motion) and
0.5 (normal diffusion). Because our trajectories extend only to
$10\,\mu$s, many distances are likely still in this diffusive regime
and have not yet reached their final equilibrium values. This
naturally shifts dynamical weight toward the longest accessible
timescales.
The effect is particularly pronounced in PDZ2L, which also exhibits
the largest discrepancy between experiment and simulation. As the
ligand becomes destabilized and gradually escapes from the binding
pocket, the associated contact distances undergo large increases
rather than small fluctuations about equilibrium, thereby strongly
enhancing the long-time dynamical response.

The comparatively weak long-time signal observed in the IR experiments
may arise from a different mechanism. IR spectroscopy is expected to
become increasingly susceptible to averaging and cancellation effects
as structural changes accumulate over longer timescales, especially in
systems such as PDZ2S and PDZ2L that undergo extensive global
reorganization. As multiple regions of the protein relax and rearrange
simultaneously, their individual spectroscopic contributions may
partially offset one another, reducing the net IR signal despite the
presence of substantial structural changes.

While we can identify several kinetic events in all of the investigated protein systems, most likely not all of them are functionally relevant; some might just be ``side-products'' of the protein response to an external stimulus. A good example in this regard is PDZ3, where we have identified six kinetic steps in experiment (and five in the MD simulations, see Fig.~\ref{fig:PDZ3}). However, we have attributed only one peak at 200~ns to the allosteric communication between the $\alpha_3$-helix and the peptide ligand, i.e., the two functionally coupled sites in this allosteric model system.\cite{bozovic_speed_2021}

As a final note we mention that the excellent agreement between
experimental and computed timescales is somewhat surprising, as
standard MD force fields are not explicitly parameterized to
accurately reproduce rare conformational transitions. Within the
contact-cluster picture, the rate-limiting step is not the breaking of a particularly strong interaction but the rare coincidence of several contact rearrangements.
In other words, the barriers governing the individual kinetic steps appear to be predominantly entropic rather than enthalpic. Consequently, the precise details of the force field parameterization may be less critical. Instead, an important factor is the topology of the configurational space, which is ultimately determined by the protein’s bonding network and is expected to be captured by a broad range of commonly used force fields. Temperature-dependent studies, both experimental and computational, could help establish the predominantly entropic origin of these barriers, and are currently underway.

%
%
\section*{Conclusions}

Time-resolved IR spectroscopy provides a uniquely direct view of protein
nonequilibrium dynamics by resolving the timescales of structural
change. 
Our central finding is that the transient IR response directly tracks
the dynamics of localized contact networks that underly
conformational
transitions. This correspondence enables the spectra to be assigned
to specific local molecular motions. 
Although the resulting logarithmic-time peak pattern of the dynamical content appears to be a universal feature of hierarchical protein relaxation, \cite{janke_universal_2025}  our analysis assigns these timescales to distinct localized contact networks and molecular rearrangements in each protein.
Mechanistically, rearrangements of
inter-residue contacts alter the protein electrostatic field, thereby
shifting the frequencies of the backbone amide I vibrations. 
Transient IR spectroscopy therefore does not merely report that
structural changes occur on particular timescales. Rather, it directly reports
the dynamics of localized contact networks that underlie conformational
transitions, providing direct experimental access to the molecular
mechanisms of protein function when combined with nonequilibrium MD
simulations.
More broadly, the combined IR/MD approach provides a general
framework for connecting experimentally observed relaxation timescales
with the molecular mechanisms that generate them.

%
%

\subsection*{Supplementary Information}

Method Section describing MD simulations, data preprocessing, and
  timescale analysis; additional
experimental data (Fig.\,\figExpPDZD)  and simulation data
(Fig.\,\figPDZDCN) of PDZ3; validations of MoSAIC
clustering (Fig.\,\figMoSAIC) and dynamical content calculations
(Fig.\,\figDynCont); performance of alternative features
(Fig.\,\figPDZDfeatures); additional simulation data of PDZ2S  (Fig.\,\figPDZZS)
and PDZ2L (Figs.\,\figPDZZL{} and \figPDZZLct); list of all peak times (Tab.\,\tablePeakTimes).

\subsection*{Acknowledgments}

The authors thank Sofia Sartore, Camilla Sordi and Steffen Wolf for
helpful comments and discussions. This work has been supported by the
Deutsche Forschungsgemeinschaft (DFG) within the framework of the
Research Unit FOR 5099 “Reducing complexity of nonequilibrium”
(project No.\ 431945604), the High Performance and Cloud Computing
Group at the Zentrum für Datenverarbeitung of the University of
Tübingen and the Rechenzentrum of the University of Freiburg, the
state of Baden-Württemberg through bwHPC and the DFG through grant
no.\ INST 37/935-1 FUGG (RV bw16I016), the ERC through grant DYNALLO,
as well as by the Swiss National Science Foundation (SNSF) through
grants 200021\_165789, 200020B\_188694, 206021\_220349.

\subsection*{Data Availability Statement}

All MD analysis tools, including the
\emph{Contacts}\cite{nagel_selecting_2023} and
\emph{MoSAIC}\cite{diez_correlation-based_2022} and \emph{TimeScaleAnalysis}
software packages, are
freely available from the Moldyn GitHub repository. Trajectories and
simulation results are available from the corresponding authors upon
reasonable request.



\providecommand{\latin}[1]{#1}
\makeatletter
\providecommand{\doi}
  {\begingroup\let\do\@makeother\dospecials
  \catcode`\{=1 \catcode`\}=2 \doi@aux}
\providecommand{\doi@aux}[1]{\endgroup\texttt{#1}}
\makeatother
\providecommand*\mcitethebibliography{\thebibliography}
\csname @ifundefined\endcsname{endmcitethebibliography}
  {\let\endmcitethebibliography\endthebibliography}{}

\end{document}